\documentclass[prl,twocolumn,showpacs]{revtex4}
\usepackage{graphicx}

\begin{document}

\title{Decoherence-Free Subspaces in Quantum Key Distribution}

\author{Zachary~D.~Walton}
\email{walton@bu.edu} \homepage[Quantum Imaging Laboratory
homepage:~]{http://www.bu.edu/qil}

\author{Ayman~F.~Abouraddy}

\author{Alexander~V.~Sergienko}

\author{Bahaa~E.~A.~Saleh}

\author{Malvin~C.~Teich}
\affiliation{Quantum Imaging Laboratory, Department of Electrical
\& Computer Engineering, Boston University, 8 Saint Mary's Street,
Boston, Massachusetts 02215-2421}


\begin{abstract}
We demonstrate that two recent innovations in the field of
practical quantum key distribution (one-way autocompensation and
passive detection) are closely related to the methods developed to
protect quantum computations from decoherence.  We present a new
scheme that combines these advantages, and propose a practical
implementation of this scheme that is feasible using existing
technology.
\end{abstract}

\maketitle

Decoherence has been a principal impediment in quantum information
processing applications.  In quantum computing,
decoherence-induced deviations from the desired computational
trajectory at the single-qubit level will quickly accumulate if
left uncorrected. Thus, techniques such as decoherence-free
subspaces (DFSs, for a review, see Ref.~\cite{Lidar03}) have been
developed as tools for protecting quantum computations.  In
quantum key distribution (QKD, for a review, see
Ref.~\cite{Gisin02}), single-qubit errors are also deleterious;
however, sufficiently infrequent single-qubit errors are
tolerable, since the resulting errors can be corrected by
classical error correction protocols. This has led many QKD
experimentalists to forego the complexity of
decoherence-mitigation techniques such as DFSs in favor of more
conventional methods to improve the precision of single-qubit
operations (periodic alignment of polarization axes, temperature
stabilization of interferometers, etc.).  In this letter, we
consider the applicability of DFSs to QKD.


This letter is organized as follows.  We begin by demonstrating
that a recently-proposed QKD implementation (one-way
autocompensating quantum cryptography~\cite{Walton02a}) is, in
fact, equivalent to a well-known DFS.  We then pursue a suggestion
in Ref.~\cite{Gisin02} to consider a single-qubit, phase-time
coding QKD scheme in which Bob is not required to actively switch
between conjugate measurement bases.  We show that both one-way
autocompensation (OWA) and passive detection are achieved by
embedding the logical Hilbert space in a larger physical Hilbert
space.  Next, we describe a new scheme that combines OWA and
passive detection. Finally, we propose an experimental
implementation of this new scheme that is feasible using existing
technology.

{\it Relating OWA and DFSs.}---In Ref.~\cite{Walton02a}, Klyshko's
``advanced wave interpretation''~\cite{Belinsky92} was used to
describe OWA as a variation on round-trip
autocompensation~\cite{Muller97,Bethune98}.  These schemes are
called autocompensating because they allows high-visibility
quantum interference without calibration or active stabilization
of the receiver's (Bob's) apparatus.  In the context of quantum
computation theory~\cite{Nielsen01}, a more natural explanation of
OWA is provided by DFSs. Palma {\it et al.}~\cite{Palma96} have
shown that a single logical qubit encoded in two physical qubits
according to
\begin{eqnarray}
|\bar{0}\rangle&\rightarrow&|01\rangle\nonumber\\
|\bar{1}\rangle&\rightarrow&|10\rangle \label{DFS}
\end{eqnarray}
will be protected against collective dephasing. Collective
dephasing describes a noise model in which each physical qubit is
subject to the same transformation
\begin{eqnarray}
|0\rangle&\rightarrow&|0\rangle\nonumber\\
|1\rangle&\rightarrow&e^{i\phi}|1\rangle\,,
\label{collective-dephasing}
\end{eqnarray}
where $\phi$ is an uncontrolled degree of freedom.  Under this
transformation, the states $|\textrm{01}\rangle$ and
$|\textrm{10}\rangle$ acquire the same phase factor ($e^{i\phi}$).
Thus, a qubit encoded according to Eq.~(\ref{DFS}) will be immune
to collective dephasing.

To link this DFS to OWA, we consider time-bin photonic
qubits~\cite{Brendel99}, in which the physical basis states
$|\textrm{0}\rangle$ and $|\textrm{1}\rangle$ correspond to early
($|\textrm{E}\rangle$) and late ($|\textrm{L}\rangle$)
single-photon wavepackets, respectively. Two-qubit states (e.g.
$|\textrm{EL}\rangle$) may be created in which the two time-bin
qubits are distinguished by some convenient degree of freedom
(e.g. polarization, or a time delay much longer than that used to
define the individual time-bin qubits themselves).

In OWA quantum cryptography, Alice superposes the two-qubit
time-bin states $|\textrm{EL}\rangle$ and $|\textrm{LE}\rangle$
with one of four relative phases ($0$, $\pi/2$, $\pi$, $3\pi/2$)
and sends the two-qubit state to Bob. Note that the superposition
of $|\textrm{EL}\rangle$ and $|\textrm{LE}\rangle$ entails
time-bin entanglement, an idea introduced in
Ref.~\cite{Brendel99}.  Bob applies one of two relative phase
shifts ($0$, $\pi/2$) to the superposed terms and makes his
measurement. In this way, they may effect the familiar four-state
QKD protocol (BB84)~\cite{Bennett84}.

\begin{figure}
\includegraphics{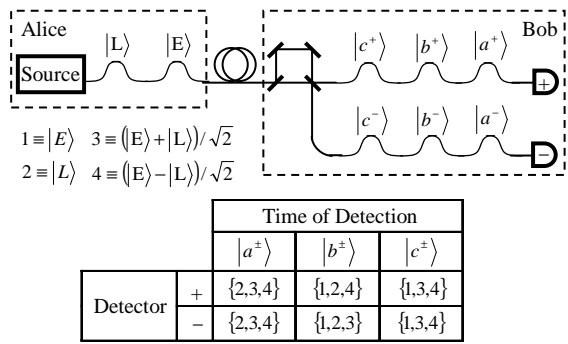}
\caption{A single-photon implementation of BB84 suggested in
Ref.~\protect\cite{Gisin02}.  The kets $|\textrm{E}\rangle$ and
$|\textrm{L}\rangle$ correspond respectively to an advanced
(early) and a delayed (late) single-photon wavepacket. Alice sends
one of the four states listed below the diagram of the apparatus.
The chart indicates which of Alice's states are consistent with a
given measurement event at Bob's side.  As described in the text,
Bob's apparatus does not require active change of measurement
basis.}\label{single-photon}
\end{figure}

The equivalence of OWA and the DFS in Eq.~(\ref{DFS}) may be seen
by carefully following Bob's detection process.  After applying
his phase shift, Bob analyzes the state using a Mach--Zehnder
interferometer (MZI) with optical delay equal to the time delay
separating $|\textrm{E}\rangle$ and $|\textrm{L}\rangle$. Using
the notation of Fig.~\ref{single-photon}, the action of the
interferometer on a single time-bin qubit is
\begin{eqnarray}
|\textrm{E}\rangle&\rightarrow&i|a^-\rangle+ie^{i\phi}|b^-\rangle-e^{i\phi}|b^+\rangle+|a^+\rangle\nonumber\\
|\textrm{L}\rangle&\rightarrow&i|b^-\rangle+ie^{i\phi}|c^-\rangle-e^{i\phi}|c^+\rangle+|b^+\rangle\,,
\label{MZ-transformation}
\end{eqnarray}
where $\phi$ is the relative phase along the two paths. Here, and
for the remainder of this letter, normalizing constants and
overall phase factors have been suppressed. By postselecting those
cases in which both photons are detected at time slots
corresponding to $|b^+\rangle$ or $|b^-\rangle$, Bob achieves the
following effective transformation:
\begin{eqnarray}
|\textrm{EL}\rangle&\rightarrow&|b^+b^+\rangle+|b^-b^-\rangle+i(|b^+b^-\rangle-|b^-b^+\rangle)\nonumber\\
|\textrm{LE}\rangle&\rightarrow&|b^+b^+\rangle+|b^-b^-\rangle-i(|b^+b^-\rangle-|b^-b^+\rangle)\,,
\label{MZ-transformation3}
\end{eqnarray}
where a common factor of $e^{i\phi}$ has no consequence.

The crucial assumption in going from Eq.~(\ref{MZ-transformation})
to Eq.~(\ref{MZ-transformation3}) is that the MZI transforms each
of the two time-bin qubits identically. For time-bin qubits
distinguished by a time delay that is short compared to the
characteristic time of interferometric drift (though long compared
to the time separating $|\textrm{E}\rangle$ and
$|\textrm{L}\rangle$), this assumption is certainly valid. The
probability of Bob detecting two photons on the same output arm
($|b^+b^+\rangle$ or $|b^-b^-\rangle$) depends on the relative
phase between the $|\textrm{EL}\rangle$ and $|\textrm{LE}\rangle$,
and similarly for the probability of detecting two photons on
different arms ($|b^+b^-\rangle$ or $|b^-b^+\rangle$).  The
critical point is that each of these probabilities is independent
of the interferometer's phase delay, $\phi$.  Thus, just as the
DFS described in Eq.~(\ref{DFS}) protects a logical qubit encoded
in two physical qubits from collective dephasing, OWA enables Bob
to measure high-visibility two-photon interference with a MZI that
does not require initial calibration or active phase
stabilization.

{\it Passive detection via enlarging the Hilbert space.}---The
two-photon quantum key distribution scheme described in
Ref.~\cite{Brendel99} has the remarkable property that both Alice
and Bob use passive detection (i.e. they are not required to
switch between conjugate measurement bases). In
Ref.~\cite{Gisin02}, Gisin et al.~suggest applying Klyshko's
advanced wave interpretation to generate an associated one-photon
scheme. We present a specific implementation of this one-photon
scheme here to show that it achieves passive detection by
enlarging the Hilbert space (see Fig.~\ref{single-photon}). Let
the advanced and delayed single-photon wavepackets be associated
with the poles of the Poincar\'{e} sphere.  The four states
required for BB84 are typically taken from the equator, since a
single MZI can be used to generate any of the equatorial states.
Instead, we imagine using two antipodal points on the equator and
the poles themselves.  Bob analyzes the signal from Alice with a
MZI, recording which detector fired (one of two possibilities) at
which time (one of three possibilities). When Bob's detection is
in the first or third time positions, he can reliably distinguish
between the pole states based on the time of detection.  When his
detection is in the second time position, he can reliably
distinguish between the equatorial states based on which detector
fired.  Thus, Bob is no longer obliged to make an active change to
his apparatus to effect the requisite change of
basis~\footnote{The idea of using pole states is explored in
Ref.~\protect\cite{Bechmann-Pasquinucci00}; however, that paper
doesn't mention the possibility of passive detection.}.

To see how this passive detection is derived from enlargement of
the Hilbert space, consider the quantum state of Alice's signal
after Bob's MZI.  Alice's four states of one qubit are mapped onto
four mutually nonorthogonal states of a six-state quantum system
(see Fig.~\ref{single-photon}). Thus, by mapping a two-state
quantum system into a six-state quantum system, Bob is able to
perform his part of the BB84 protocol with a fixed-basis
measurement in the six-state Hilbert space~\footnote{A similar
idea is presented in Ref.~\protect\cite{Inoue02}.  In that paper,
Alice uses four states of a three-state quantum system, and Bob
achieves passive detection by mapping Alice's three-state quantum
system into an eight-state quantum system.}.

\begin{figure*}
\includegraphics{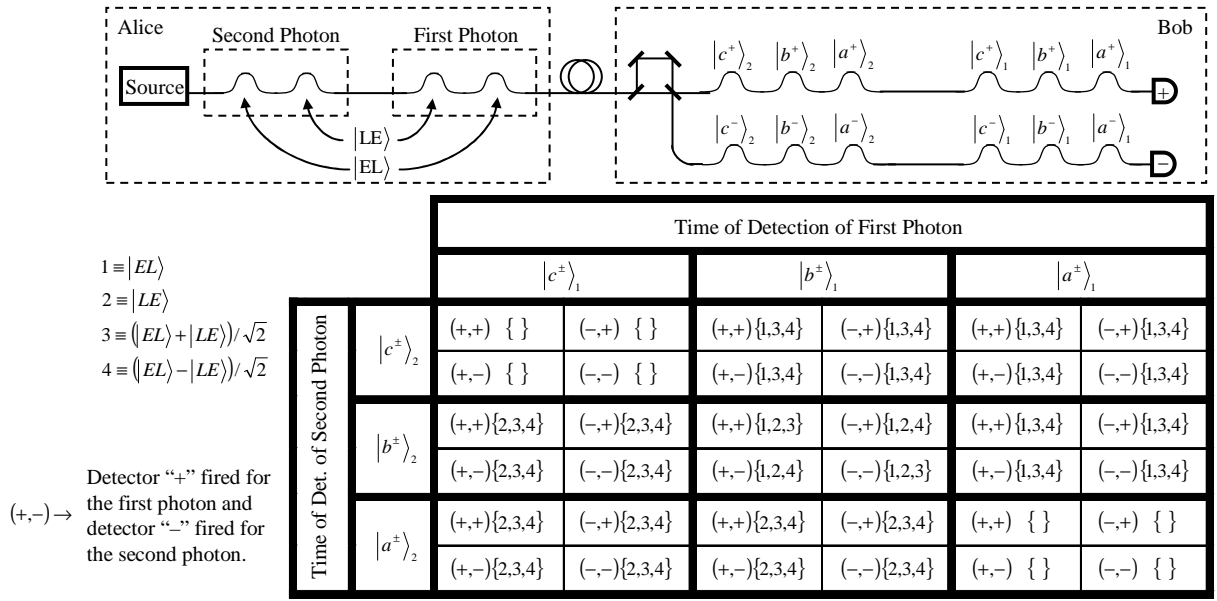}
\caption{A new scheme for quantum key distribution that combines
OWA with passive detection.  Two time-bin qubits are sent from
Alice to Bob in one of the four quantum states on the left of the
figure.  The chart on the right uses two levels of structure to
describe the detection pattern at Bob's side.  The coarse
structure is defined by the bold lines. Each of the nine
bold-lined rectangles corresponds to a specification of the joint
time of detection of the two photons. The fine structure is
defined by the thin lines.  Each of the four thin-lined rectangles
within a bold-lined rectangles corresponds to a specification of
which detector fired for each of the two photons (this coding is
illustrated by an example at the bottom left of the figure).  The
numbers in the curly brackets in each thin-lined rectangle
indicates which (if any) of the four quantum states on the left
are consistent with the corresponding detection
pattern.}\label{two-photon}
\end{figure*}

{\it Combining OWA and passive detection.}---OWA and passive
detection have been previously presented in separate proposals
(Refs.~\cite{Walton02a} and~\cite{Brendel99}, respectively).  Here
we present a new scheme that combines these two beneficial
features in a single implementation (see Fig.~\ref{two-photon}).
The new scheme follows from that presented in
Ref.~\cite{Walton02a}, just as the preceding single-photon scheme
follows from the traditional phase-coding implementation. Let the
states $|1\rangle$ and $|2\rangle$ in Fig.~\ref{two-photon} be
associated with the poles of the Poincar\'{e} sphere.  Instead of
using equatorial states and forcing Bob to postselect those cases
for which the advanced (delayed) amplitudes take the long (short)
path, we use two equatorial points ($|3\rangle$ and $|4\rangle$)
and the poles themselves to make up Alice's four signal states.
Signal states that are consistent with a given joint detection are
presented in the chart. As seen in Fig.~\ref{single-photon}, each
photon can lead to six different detection events.  Thus, since
the new protocol involves two photons, there are 36 possible
detection events (see Fig.~\ref{two-photon}).

The protocol operates as follows.  As in BB84, Alice and Bob
publicly agree on an association of each of the four signal states
(see Fig.~\ref{two-photon}) with logical values ``0'' or ``1''
(i.e., $1\rightarrow$ ``0'', $2\rightarrow$ ``1'', $3\rightarrow$
``0'', $4\rightarrow$ ``1''). For each run of the experiment,
Alice randomly chooses one of the four signal states and sends it
to Bob. When Bob detects both photons in their respective middle
time slots, he has effectively measured in the $\{3,4\}$ basis
(the ``phase'' basis). When Bob detects both photons in their
early time slots, or both photons in their late time slots, he has
effectively measured in the $\{1,2\}$ basis (the ``time''
basis)~\footnote{On the occasions when Bob's detection pattern is
(early,~middle), (middle,~early), (middle,~late), or
(late,~middle), he has also effectively measured in the time
basis. However, to simplify the analysis by making the probability
of successful bit-sharing independent of the basis in which Alice
sent, we only consider the extreme cases (early,~early) and
(late,~late) as valid time-basis detections.}. After the quantum
transmission, Alice and Bob publicly announce their bases. On the
occasions when their bases match, Bob is able to infer the state
that Alice sent, based on his detection pattern using the chart in
Fig.~\ref{two-photon}. As in single-qubit BB84, the occasions in
which their bases do not match are discarded. The scheme achieves
passive detection (Bob is not required to make any active changes
to his apparatus) and autocompensation (the phase delay in Bob's
interferometer does not affect any measured probabilities).  The
intrinsic efficiency of the scheme is $1/4$, compared to $1/2$ for
single-qubit BB84.

\begin{figure}
\includegraphics{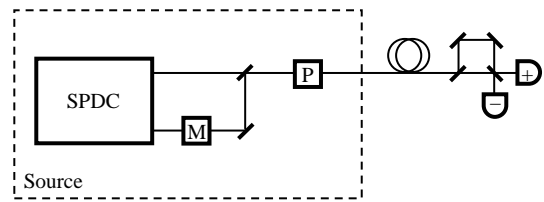}
\caption{A proposed implementation for the source employed in
Fig.~\protect\ref{two-photon}. ``SPDC'' is a nonlinear crystal
pumped by a brief pulse to produce a noncollinear,
polarization-entangled two-photon state via spontaneous parametric
down-conversion. The action of elements ``M'' and ``P'' is
described in the text.}\label{source}
\end{figure}

{\it A feasible implementation.}---A proposed implementation for
the source employed in Fig.~\ref{two-photon} is presented in
Fig.~\ref{source}.  First, a pair of noncollinear,
polarization-entangled photons is produced via type-II spontaneous
parametric down-conversion from a nonlinear crystal pumped by a
brief pulse~\footnote{A femtosecond pump pulse is typically
desired for experiments involving the simultaneous creation of
multiple down-converted photon pairs~\protect\cite{DeMartini01}.
Our implementation does not require such a brief pump pulse, and
will work with a picosecond laser, such as that used in
Ref.~\protect\cite{Tittel00}.}. Second, the modulating element
``M'' performs one of four functions (filter one of the two
polarization modes, or introduce one of two relative phases
between the two polarization modes), based on Alice's choice of
signal states. Third, the two beams are combined with a relative
temporal delay that matches the temporal delay Bob will
subsequently introduce with his MZI. This stage converts the
photon pair from a pair of spatially-defined
polarization-entangled qubits to a pair of polarization-defined
time-bin entangled qubits. Finally, the element labeled ``P'' (for
polarization) delays and rotates one of the polarization modes by
a duration much greater than the delay of the third step, such
that the delayed portion of the state in the same polarization as
the non-delayed portion. Thus, the two photons sent from Alice to
Bob have the wavepacket structure illustrated at the top of
Fig.~\ref{two-photon}.

There are two noteworthy aspects of the configuration in
Fig.~\ref{source}.  First, the technique introduced in
Ref.~\cite{Brendel99} for creating time-bin entangled photons
pairs only leads to superpositions of the correlated possibilities
(i.e. $|\textrm{EE}\rangle$ and $|\textrm{LL}\rangle$). The source
presented in Fig.~\ref{source} enables arbitrary superpositions of
the anti-correlated possibilities (i.e. $|\textrm{EL}\rangle$ and
$|\textrm{LE}\rangle$). Furthermore, the correlated states can
easily be created from this source by rotating the polarization
axes at element ``M'' in Fig.~\ref{source}.  In this way, all four
time-bin entangled Bell states can be conveniently generated with
this source. Second, the interference in Bob's interferometer
results from the indistinguishability of photon amplitudes that
were initially in the same polarization mode.  This is in contrast
to configurations in which photon amplitudes from different
polarization modes are made indistinguishable by use of a
polarization analyzer.  Thus, the reduction in visibility that has
come to be associated with extremely brief pump
pulses~\cite{Keller97} will not be present in this scheme. Note
that a symmetrization method has been developed to restore
visibility for experiments using polarization-entangled photons
created by such a short pulse pump~\cite{DeMartini01,Kim02}.

{\it Conclusion.}---We have demonstrated that two recent
innovations in the field of practical quantum key distribution
(autocompensation and passive detection) are closely related to
the methods developed to protect quantum computations from
decoherence. Pursuing this conceptual link between techniques from
quantum computation and advances in practical QKD, we have
developed a new QKD scheme (Fig.~\ref{two-photon}) that combines
autocompensation and passive detection.  Furthermore, we have
proposed a practical implementation of the scheme
(Fig.~\ref{source}) that is feasible using existing technology.

We thank Matthew D. Shaw and Magued B. Nasr for valuable
conversations, and Jean C. Boileau for constructive criticism on
an early draft.  This work was supported by the National Science
Foundation; the Center for Subsurface Sensing and Imaging Systems
(CenSSIS), an NSF Engineering Research Center; and the Defense
Advanced Research Projects Agency (DARPA).


\end{document}